\documentstyle[12pt,epsf]{article}

\catcode`\@=11

\def\section{\@startsection{section}{1}{\z@}
  {3.5ex plus 1.0ex minus 0.2ex}{2.3ex plus .2ex}{\normalsize}}
\def\subsection{\@startsection{subsection}{2}{\z@}
  {3.25ex plus 1.0ex minus 0.2ex}{1.5ex plus 0.2ex}{\normalsize\bf}}
\def\subsubsection{\@startsection{subsubsection}{3}{\z@}
  {3.25ex plus 1.0ex minus 0.2ex}{1.5ex plus 0.2ex}{\normalsize\bf}}

\@addtoreset{equation}{section}

\catcode`\@=12

\newcommand{\skipline}{\vspace{\baselineskip}}

\def\fun#1#2{\lower3.6pt\vbox{\baselineskip0pt\lineskip.9pt
  \ialign{$\mathsurround=0pt#1\hfil##\hfil$\crcr#2\crcr\sim\crcr}}}

\begin{document}

\begin{titlepage}

\begin{center}
GLUEBALL MASS PREDICTIONS OF THE VALENCE APPROXIMATION\\
TO LATTICE QCD
\end{center}

\skipline
\skipline

\begin{center}
H. Chen, 
J. Sexton\footnote{permanent address: Department of Mathematics,
Trinity College, Dublin 2,
Republic of Ireland}, 
A. Vaccarino,\\
and D. Weingarten \\

IBM Research \\
P.O. Box 218, Yorktown Heights, NY 10598
\end{center}

\skipline
\skipline

\begin{center}
ABSTRACT
\end{center}

\begin{quotation}

We evaluate the infinite volume, continuum limit of glueball masses in
the valence (quenched) approximation to lattice QCD. For the lightest
states with $J^{PC}$ of $0^{++}$ and $2^{++}$, we obtain $m_0 = 1340
\pm 160$ MeV and $m_2 = 1900 \pm 320$ MeV.

\end{quotation}

\skipline
\skipline

\end{titlepage}

\section{INTRODUCTION}\label{intro}

Although dozens of observed hadrons have been convincingly interpreted
as bound states of combinations of quarks and antiquarks, physical
particles composed primarily of chromoelectric field, glueballs, have so
far not been identified unambiguously in experiment.  One possible
explanation is that glueballs simply may not exist. All excitations of
the chromoelectric field might occur only mixed with a large number of
quark-antiquark states. Each mixed state might have only a small
probability amplitude of excited chromoelectric field and, as a result,
show no easily identifiable experimental signature of the presence of a
field excitation.  Alternatively, some of the particles already observed
in experiment may actually be glueballs and may not have been identified
because QCD's predictions for glueball masses and decays have not yet
been determined accurately enough.

In the present article, we evaluate the infinite volume, continuum limit
of the predictions of lattice QCD, in the valence approximation, for the
masses of the lightest glueballs with $J^{PC}$ of $0^{++}$ and $2^{++}$.
The $0^{++}$ is probably the lightest member of the glueball spectrum
and the $2^{++}$ the next lightest \cite{Berg,Michael88}.  The valence
approximation may be viewed as replacing the momentum dependent color
dielectric constant arising from quark-antiquark vacuum polarization
with its zero-momentum limit \cite{Weingar81}.  For glueball masses,
this approximation amounts to a reinterpretation of the predictions of
pure gauge theory. Pure gauge QCD certainly predicts the existence of
unmixed glueballs.

The valence approximation could be expected to be fairly reliable for
low-lying hadron masses, which are determined largely by the low
momentum behavior of the chromoelectric field. This expectation is
supported by a recent valence approximation calculation~\cite{Butler93}
of the masses of eight low-lying baryons and quark-antiquark mesons.
The predicted masses are all within 6\% of experiment and all
differences between prediction and experiment are consistent with the
calculation's statistical uncertainties.  The prediction we obtain here
for the $0^{++}$ glueball mass, $1340 \pm 160$ MeV, lies below the
largest mass, $m_{\Omega} = 1672$ MeV, for which the valence
approximation has now, to some degree, been established as reliable.
Allowing a disagreement of up to $80$ MeV, 6\% of $1340$ MeV, between
the valence approximation and full QCD, the predicted $0^{++}$ mass is
consistent with the masses of either of two well-established $0^{++}$
states, $f^0(1400)$ and $f^0(1590)$.  The mass prediction for the
$2^{++}$ glueball, $1900 \pm 320$ MeV, is close enough to the range for
which the valence approximation has been tested that it appears
reasonable to expect a disagreement of at most 6\% with full QCD in this
case also, giving an additional uncertainty of $120$ MeV.  The overall
uncertainty in the $2^{++}$ mass is large enough that this state could
be any one of a half dozen or so observed resonances.

Infinite volume, continuum limit glueball masses obtained from the
valence approximation to lattice QCD have now also been reported in
Ref.~\cite{Bali93}, which appeared while the present work was in
progress. The predicted $0^{++}$ and $2^{++}$ masses are $1550 \pm 50$
MeV and $2270 \pm 100$ MeV, respectively. The prediction for the
$0^{++}$ mass is about 1.3 standard deviations above our result.  It
seems to us that the central values of both predictions may be somewhat
larger and the error bars of both predictions somewhat smaller than the
data reported in Ref.~\cite{Bali93} actually suggests.  We will return
to this subject below.

The calculations described here were done on the GF11 parallel computer
at IBM Research \cite{Weingar90} and took about six months to complete.
GF11 was used in configurations ranging from 375 to 480 processors, with
sustained speeds ranging from 5 Gflops to 7 Gflops.

\section{DEFINITIONS AND ALGORITHMS}

To define propagators from which light glueball masses can be extracted
efficiently, it is useful to form smeared gauge fields with reduced
coupling to the underlying gauge field's high momentum
components \cite{Grif,Tep}. In place of the gauge invariant smearing of
Refs.~\cite{Grif,Tep}, however, we use a smearing method based on
Coulomb gauge.  Each gauge configuration on which glueball propagators
are to be constructed is transformed to lattice Coulomb gauge, defined
to maximize at each site the target function $\sum_{x i} Re Tr [ U_i(
x)]$, where the sum is over all lattice sites and space direction gauge
links.  A gauge transformation which produces a local maximum of this
sum is found by a method qualitatively similar to the
Cabbibo-Marinari-Okawa Monte Carlo algorithm. The lattice is swept
repeatedly, and at each site we maximize the target function first by a
gauge transformation in the $SU(2)$ subgroup of $SU(3)$ acting only on
gauge index values 1 and 2, then by a gauge transformation in the
$SU(2)$ subgroup acting only on index values 2 and 3, then by a gauge
transformation in the subgroup acting only on index values 1 and 3.
Maximizing the target function over $SU(2)$ subgroups is easier to
program than a direct maximization over all of $SU(3)$.  On the other
hand, it is not clear that maximizing each site over $SU(3)$ would
significantly accelerate the full transformation to Coulomb gauge.  A
local maximum is reached when at each site the quantity $R( x)$ vanishes
where
\begin{eqnarray}
\label{defqr}
Q( x) & = & \sum_i [ U_i( x) - U_i^{\dagger}( x - \hat{i}) ], \nonumber \\
R( x) & = & Q( x) - Q^{\dagger}( x) - \frac{2}{3} Im Tr[ Q( x)]. 
\end{eqnarray}
The vector $\hat{i}$ is a unit lattice vector in the positive $i$
direction.  We stop the iteration process when the sum over the lattice
of the quantity $Tr [ R^{\dagger}( x) R( x)] $ becames smaller than a
convergence parameter $r$.

Coulomb gauge smeared link variables $U_i^s( x)$ for $s \ge 1$ are then
defined by the average
\begin{eqnarray}
\label{defu}
U_i^s( x) = \frac{1}{(s + 1)^2} \sum_{0 \le p, q \le s} U_i( x + p
\hat{j} + q \hat{k}),
\end{eqnarray}
where $j$ and $k$ are the two space directions orthogonal to $i$.
The product of $s$ sequential $U_i^s( y)$ gives $V_i^s( x)$,
\begin{eqnarray}
\label{defv}
V_i^s( x) = U_i^s( x) \ldots  U_i^s[ x + (s - 1) \hat{i}].
\end{eqnarray}
From the $V_i^s( y)$ we define smeared loops $L_{i j}^s( x)$ 
\begin{eqnarray}
\label{defl}
L_{i j}^s( x)  =  Tr[ V_i^s( x + s \hat{i}) V_j^s( x + 2 s \hat{i} + s
\hat{ j}) V_i^s( x + s \hat{i} + 2 s \hat{ j})^{\dagger} 
V_j^s( x + s \hat{ j})^{\dagger} ].
\end{eqnarray}
This definition fits together as smoothly as possible the products
of smeared links forming each side of $L_{i j}^s( x)$.  The sum of
$L_{i j}^s( x)$ over all $x$ in a hyperplane with fixed time component
$t$ defines the zero-momentum loop variable $\bar{L}_{ij}^s( t)$.

For smearing size 0, the loop variable
$L_{i j}^0( x)$ is defined to be an unsmeared plaquette
\begin{eqnarray}
\label{defl0}
L_{i j}^0( x)  = Tr [ U_i( x ) U_j( x + \hat{i}) 
U_i( x + \hat{ j})^{\dagger} U_j( x)^{\dagger} ], 
\end{eqnarray}
and $\bar{L}_{i j}^0( x)$ is then defined from $L_{i j}^0( x)$.

A field for the ${0^{++}}$ glueball, transforming according to the $A_1$
representation of the cubic symmetry group
\cite{Berg}, is given by 
\begin{eqnarray}
\label{defa}
A^s( t) = \sum_{ij} Re [ \bar{L}_{ij}^s( t)].
\end{eqnarray}
Fields for the ${2^{++}}$ glueball, transforming according to the
$E$ representation of the cubic symmetry group, are
given by 
\begin{eqnarray}
\label{defe}
E_1^s( t) & = & Re [ \bar{L}_{12}^s( t) - \bar{L}_{23}^s( t) ],
\nonumber \\ 
E_2^s( t) & = & Re [ \bar{L}_{12}^s( t) + \bar{L}_{23}^s(t) - 2
\bar{L}_{13}^s( t) ]/\sqrt{3}   
\end{eqnarray}
From these glueball fields, we define glueball propagators as
\begin{eqnarray}
\label{defc}
C_0^{ss'}(t) & = & < A^s(t) A^{s'}(0) > - < A^s(t)>< A^{s'}(0)>,
\nonumber \\
C_2^{ss'}(t) & = & 
\sum_i [ < E_i^s(t) E_i^{s'}(0) > - < E_i^s(t) > < E_i^{s'}(0) >] ,
\end{eqnarray}
Although $< E_i^s(t) >$ vanishes for a lattice which is invarient under
the cubic symmetry group acting on the space directions, one of the
lattices we use is slightly asymmetric, $30 \times 32^2$. This lattice
is close enough to symmetric, however, that the contribution of the
disconnected term in $C_2^{ss'}(t)$ turn out to be significantly smaller
than our statistical errors.

For any pair of sizes $s$ and $s'$, at sufficiently large values of $t$
and the lattice time direction periodicity $T$, the glueball propagators
approach the asymptotic form
\begin{eqnarray}
\label{defasym}
C_i^{ss'}(t)  \rightarrow  Z_i^{ss'} \{ exp( -m_i t) + exp[ -m_i (T -t)]\}
\end{eqnarray}
where $m_0$ is the energy eigenvalue
of the lightest $0^{++}$ zero-momentum state $|0^{++}>$, 
$m_2$ is the common energy eigenvalue
of the lightest $2^{++}$ zero-momentum states $|2^{++}, i>$ 
with spin index $i = 1, 2$, and $Z_0^{ss'}$ and $Z_2^{ss'}$ are related
to the vacuum state $| \Omega >$ matrix elements
\begin{eqnarray}
\label{defz}
Z_0^{ss'} & = & < \Omega |  A^s(t) | 0^{++}  > < 0^{++} |  A^{s'}(t)
| \Omega >, \nonumber \\
Z_2^{ss'} & = & 
\sum_i < \Omega |  E_i^s(t) | 2^{++}, i  > 
< 2^{++}, i |  E_i^{s'}(t) | \Omega >.
\end{eqnarray}

It is not hard to show that, for any value of $t$, the propagators
$C_i^{ss'}(t)$ are sums of terms of the form in Eqs.~(\ref{defasym})
with each $Z_0^{ss'}$ matrix hermitian and non-negative.  Thus each
propagator with the same size smearing at both ends, $C_i^{ss}(t)$, is a
convex function.  To extract a glueball mass from each propagator
$C_i^{ss'}(t)$, we search for a range of $s$, $s'$ and $t$ for which the
asymptotic form is approached. We then fit the asymptotic form to this
data set, minimizing the fit's $\chi^2$ determined from the full
correlation matrix of the target data set.

Statistical uncertainties are found for the resulting masses by the
bootstrap method \cite{Efron}.  From each ensemble of N gauge field
configurations, we generate a collection of 100 bootstrap ensembles.
Each bootstrap ensemble consists of N configurations selected randomly
from the underlying N member ensemble, allowing repeats. For each of the
bootstrap ensembles, masses are determined by repeating our fitting
method from the beginning but using the fitting interval already chosen
for the full ensemble. The statistical uncertainty of a mass prediction
is taken to be half the difference between a mass which is higher than
all but 15.9\% of the bootstrap masses and a mass which is lower than
all but 15.9\% of the bootstrap masses. In the limit of large N, for
which the collection of bootstrap masses will approach a gaussian
distribution, the definition we use for statistical uncertainty
approaches the dispersion, $d$, given by $d^2 = < m^2 > - < m >^2$.

Our method of determining masses and error bars is nearly the same as
used in Ref.~\cite{Butler93}.  The sole difference is that in the
calculation of Ref.~\cite{Butler93} the fitted range of $s$, $s'$ and
$t$ is chosen independently on each different bootstrap data set to
minimize the fit's value of $\chi^2$ per degree of freedom. Although
allowing the fitted range of $s$, $s'$ and $t$ to be chosen
independently for each bootstrap ensemble appears to be logically the
best procedure, we were forced to use a fixed fitting range in the
present calculation.  The statistical fluctuations in glueball
propagators for some bootstrap data sets were sufficiently large that an
automated choice of fitting range would be difficult to implement, while
choosing fitting ranges by hand for each bootstrap ensemble would be too
time consuming.

It is perhaps worth mentioning that the values of glueball masses
obtained from Eqs.~(\ref{defasym}) are independent of the choice of the
convergence parameter $r$.  Coulomb gauge leaves the time direction link
matrices unconstrainted.  The integration over time direction link
matrices in the path integral generates, in the corresponding time
evolution transfer matrix of the Hilbert space formulation of lattice
QCD \cite{Luscher77}, a projection operator onto the gauge invariant
sector of Hilbert space.  For values of $r$ which are large, the states
created by our glueball operators will include large gauge variant
components. These are canceled off by the time evolution projection
operators to the gauge invariant sector of Hilbert space.  The resulting
propagators of Eqs.~(\ref{defc}) exhibit only the masses of physical,
gauge invariant states.  The price of this cancellation is an increase
in the statistical errors in masses. In practice we choose a small
enough value of $r$ so that this increase in statistical errors is
negligible.

In comparison to the gauge invarient smearing methods of
Refs.~\cite{Grif,Tep}, the numerical cost of fixing to Coulomb gauge is
rather large. As a results our method of constructing glueball
propagators is less efficient that those of Refs.~\cite{Grif,Tep}.  An
advantage of our procedure is that it does a more complete job of
eliminating high frequency components of the gauge field than do the
methods of Refs.~\cite{Grif,Tep} and, correspondingly, is free of the
potential problems discussed in Ref.~\cite{Kron}. A comparison of our
data with that of Ref.~\cite{Bali93} also suggests that, for a fixed
ensemble size, our method may give somewhat smaller statistical errors
in glueball propagators at large time separations.

Our procedure for determining glueball masses from glueball propagators
appears to be marginally more reliable than two competing methods.  We
found that fitting the full propagator matrix $C_i^{ss'}(t)$ gives
masses which are somewhat more stable to variations in ensemble size
than either fitting the correlation function $C_i^{ss}(t)$ for a single
optimal smearing size $s$, or fitting the correlation function
$\sum_{ss'} \zeta^s \zeta^{s'} C_i^{ss'}(t)$ for a single optimal vector
$\zeta^s$ \cite{Michael88}. It also appears to us that a single fit to
the full set of data provides the least biased way of resolving the
small disagreements which occur among the masses found with various
different combinations of smearing size for the propagator's source and
sink.

\section{MASSES}

Gauge configurations were generated according to the standard Wilson one
plaquette action using the Cabbibo-Marinari-Okawa algorithm.
Table~\ref{tab:lattices} lists the lattice sizes, $\beta = 6 / g_0^2$
values, sweeps skipped between gauge configurations, and number of
configurations used in the ensembles from which glueball propagators
were calculated. Each configuration was equilibrated for at least 30000
sweeps before we began collecting data.  The number of sweeps between
configurations on which data was collected, however, was not large
enough for us to be certain whether or not successive data samples were
correlated.  We therefore checked our bootstrap evaluation of error bars
to be sure that it was not distorted by the presence of correlations. At
$\beta$ of 5.7, 6.17 and 6.4, our collections of gauge configurations
were divided into sequential groups of 2, 4, 8 and 16 gauge
configurations.  Glueball propagators were averaged over each bin, and a
bootstrap error evaluation was done on masses determined from the binned
ensembles.  Any correlations present in an ensemble of propagators
before binning become progressively weaker in the ensembles with larger
bins. Our error estimates showed no significant variation as bins were
made larger.  A direct calculation of the correlation between successive
glueball propagators in the ensembles of Table~\ref{tab:lattices} also
produced no statistically significant evidence for correlations.

For all the lattices listed in Table~\ref{tab:lattices}, the gauge
fixing convergence parameter $r$ was set to $10^{-5}$.  The average
number of sweeps required for convergence to this accuracy ranged from
1525 on the lattice $24^3 \times 36$ at $\beta$ of 5.93, to 2270 on the
lattice $16^3 \times 32$ at $\beta$ of 5.7.

\begin{figure}
\epsfxsize=\textwidth \epsfbox{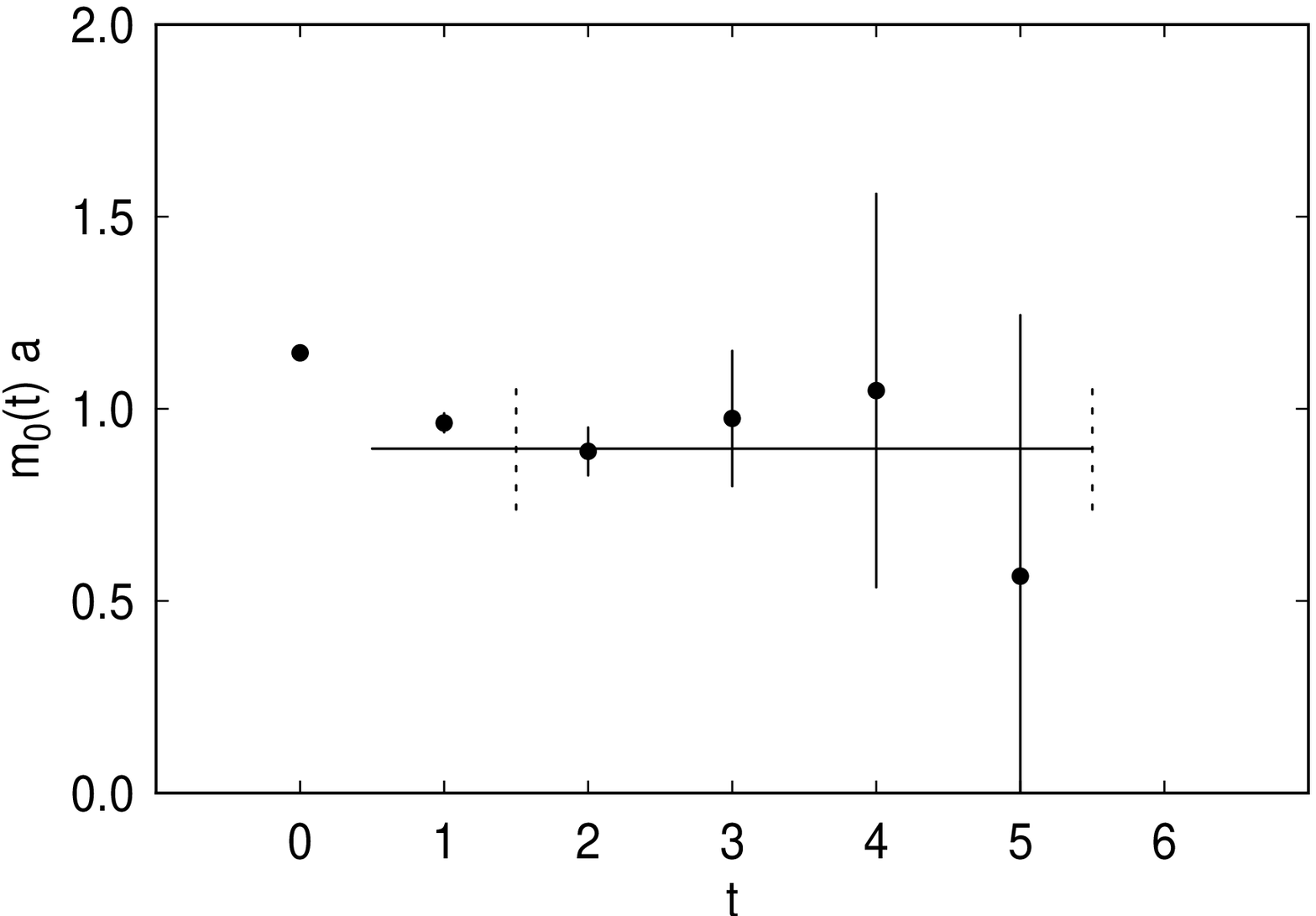}
\caption{ Effective masses and fitted mass for the $0^{++}$ glueball on a
$16^3 \times 24$ lattice at $\beta = 5.7$ for smearing size 2.}
\label{fig:m0at57}
\end{figure}

\begin{figure}
\epsfxsize=\textwidth \epsfbox{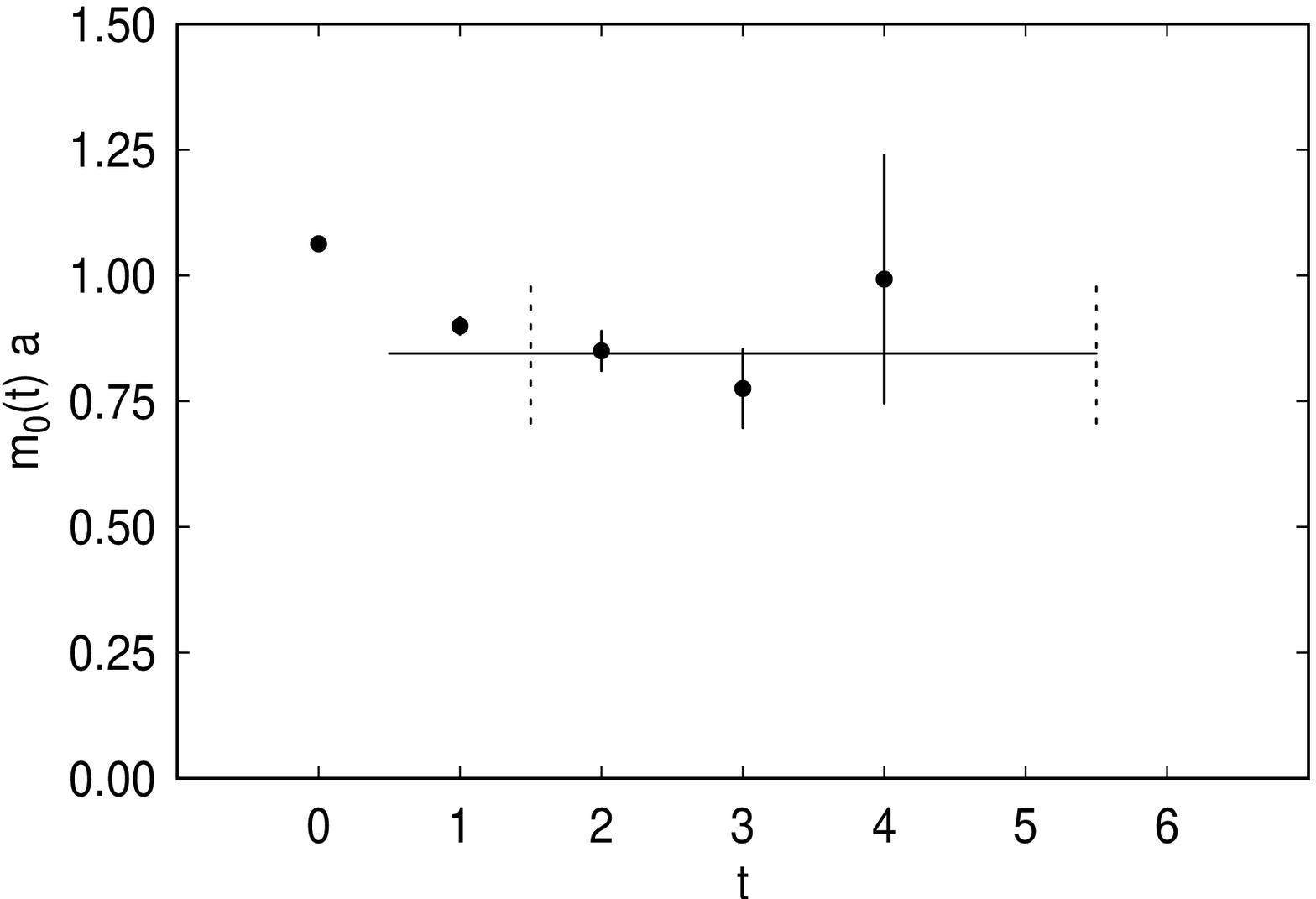}
\caption{ Effective masses and fitted mass for the $0^{++}$ glueball on a
$20^3 \times 30$ lattice at $\beta = 5.83$ for smearing size 2.} 
\label{fig:m0at58}
\end{figure}

\begin{figure}
\epsfxsize=\textwidth \epsfbox{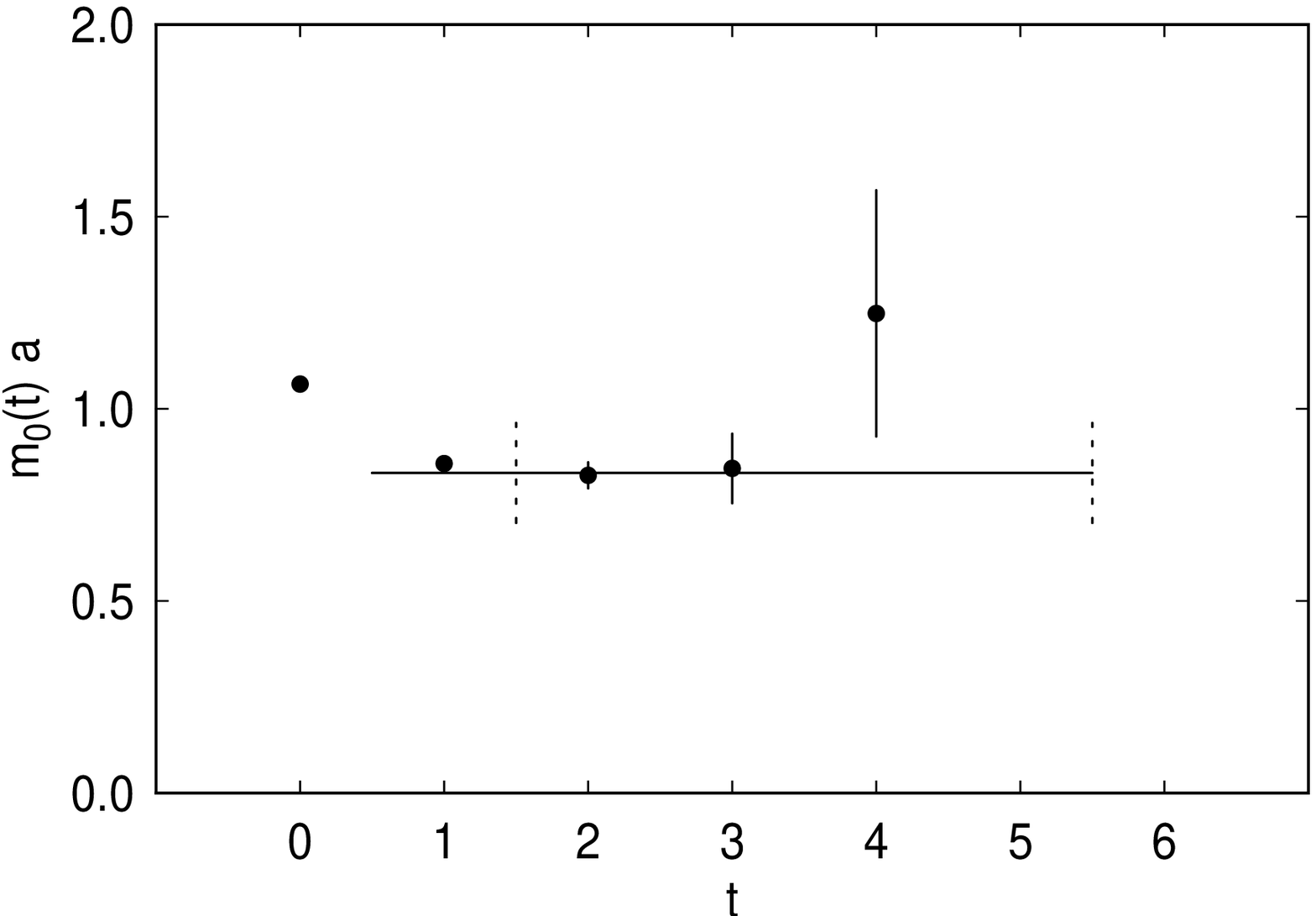}
\caption{ Effective masses and fitted mass for the $0^{++}$ glueball on a
$24^3 \times 36$ lattice at $\beta = 5.93$ for smearing size 2.}
\label{fig:m0at59}
\end{figure}

\begin{figure}
\epsfxsize=\textwidth \epsfbox{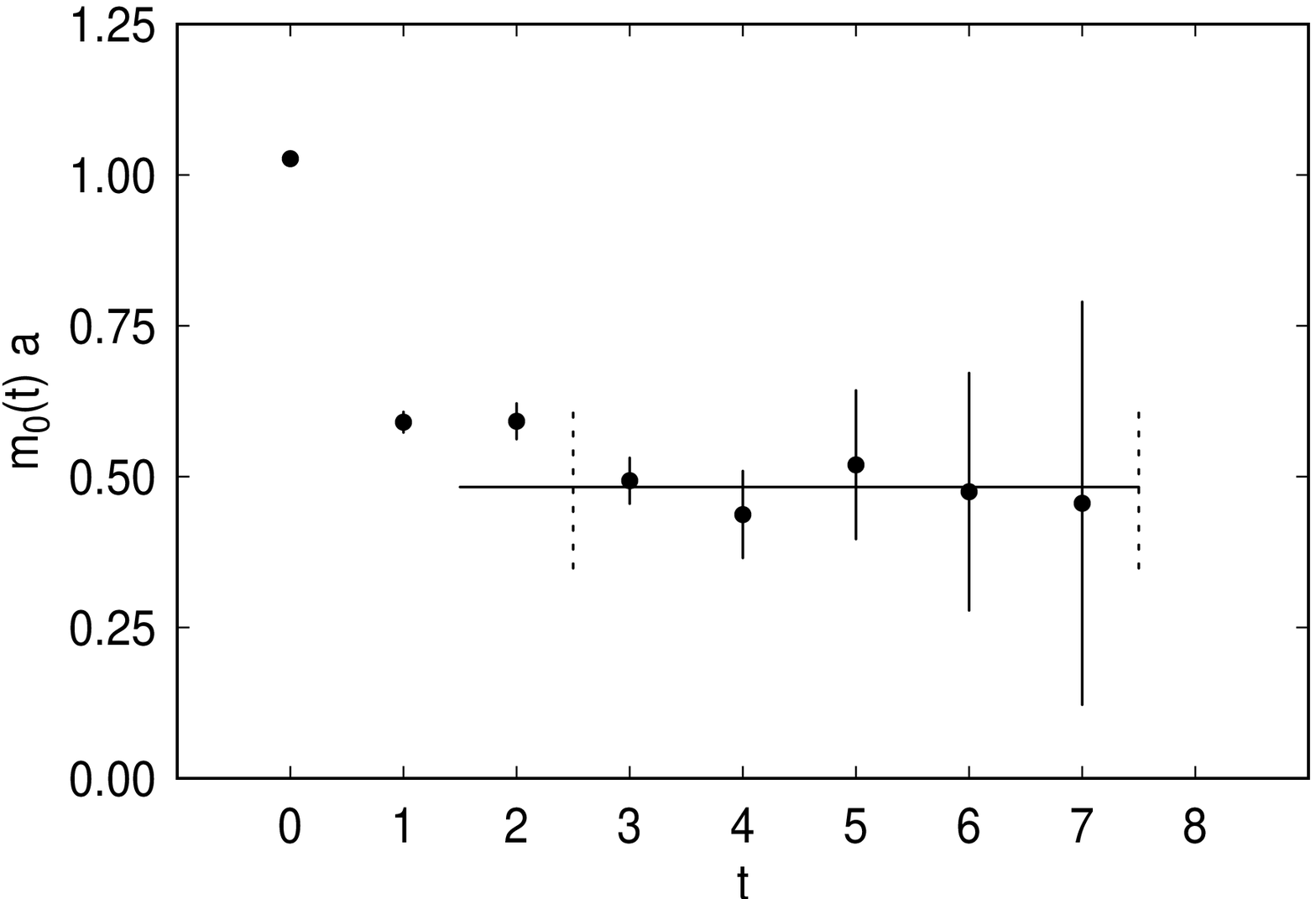}
\caption{ Effective masses and fitted mass for the $0^{++}$ glueball
with smearing size 4 on a
$30 \times 32^2 \times 40$ lattice at $\beta = 6.17$.}
\label{fig:m0at61}
\end{figure}

\begin{figure}
\epsfxsize=\textwidth \epsfbox{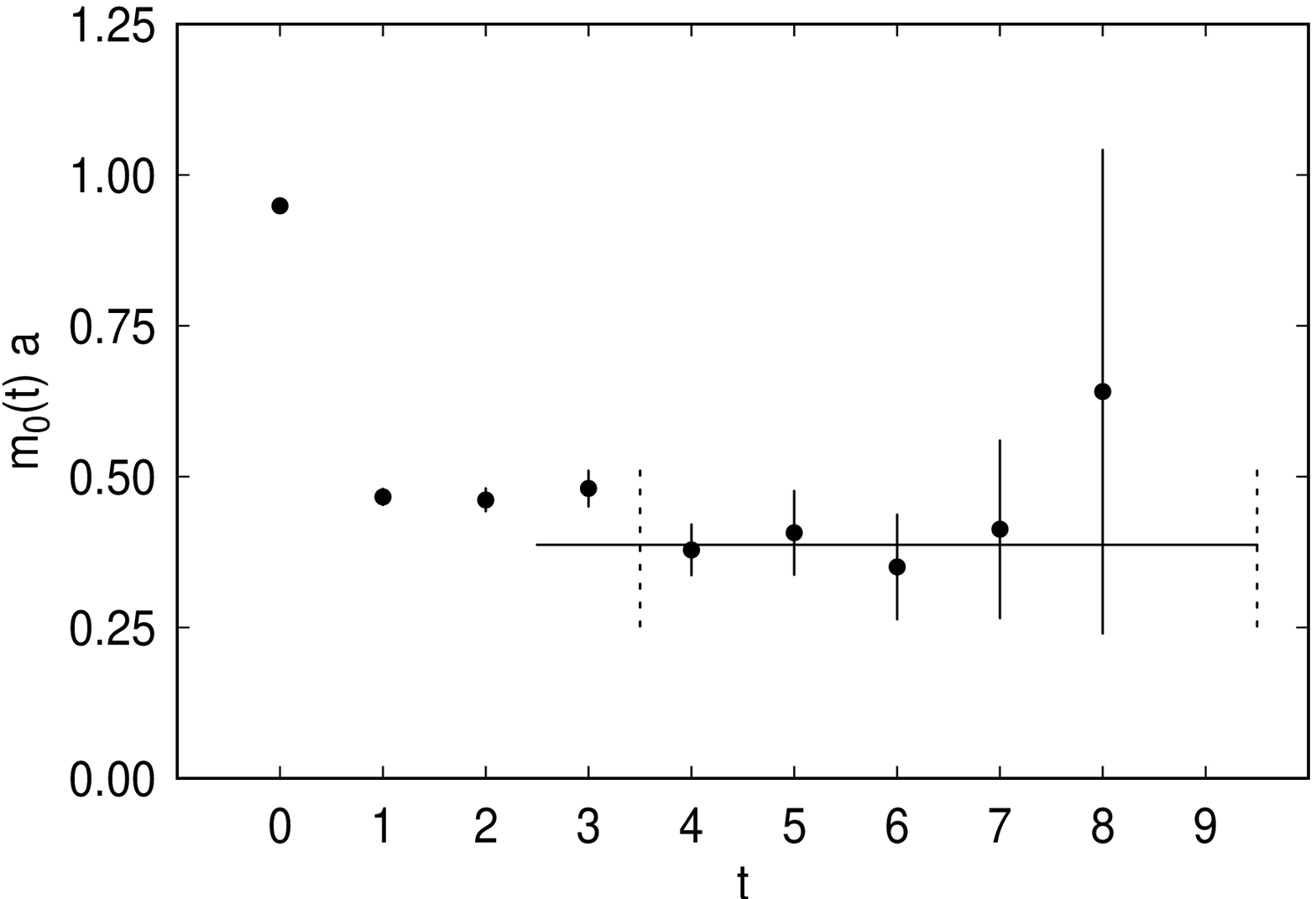}
\caption{ Effective masses and fitted mass for the $0^{++}$ glueball with
smearing size 5 on a
$30 \times 32^2 \times 40$ lattice at $\beta = 6.40$.}
\label{fig:m0at64}
\end{figure}

\begin{figure}
\epsfxsize=\textwidth \epsfbox{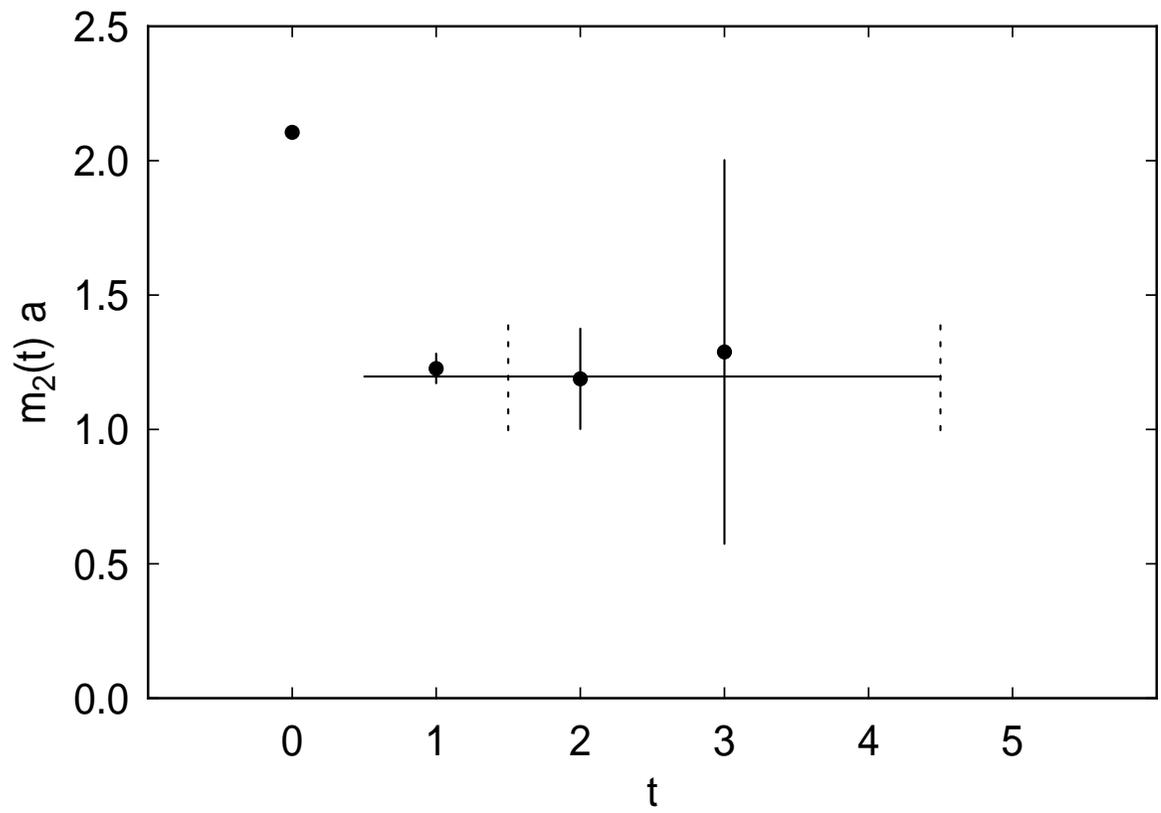}
\caption{ Effective masses and fitted mass for the $2^{++}$ glueball 
with smearing size 4 on a
$24^3 \times 36$ lattice at $\beta = 5.93$.}
\label{fig:m2at59}
\end{figure}

\begin{figure}
\epsfxsize=\textwidth \epsfbox{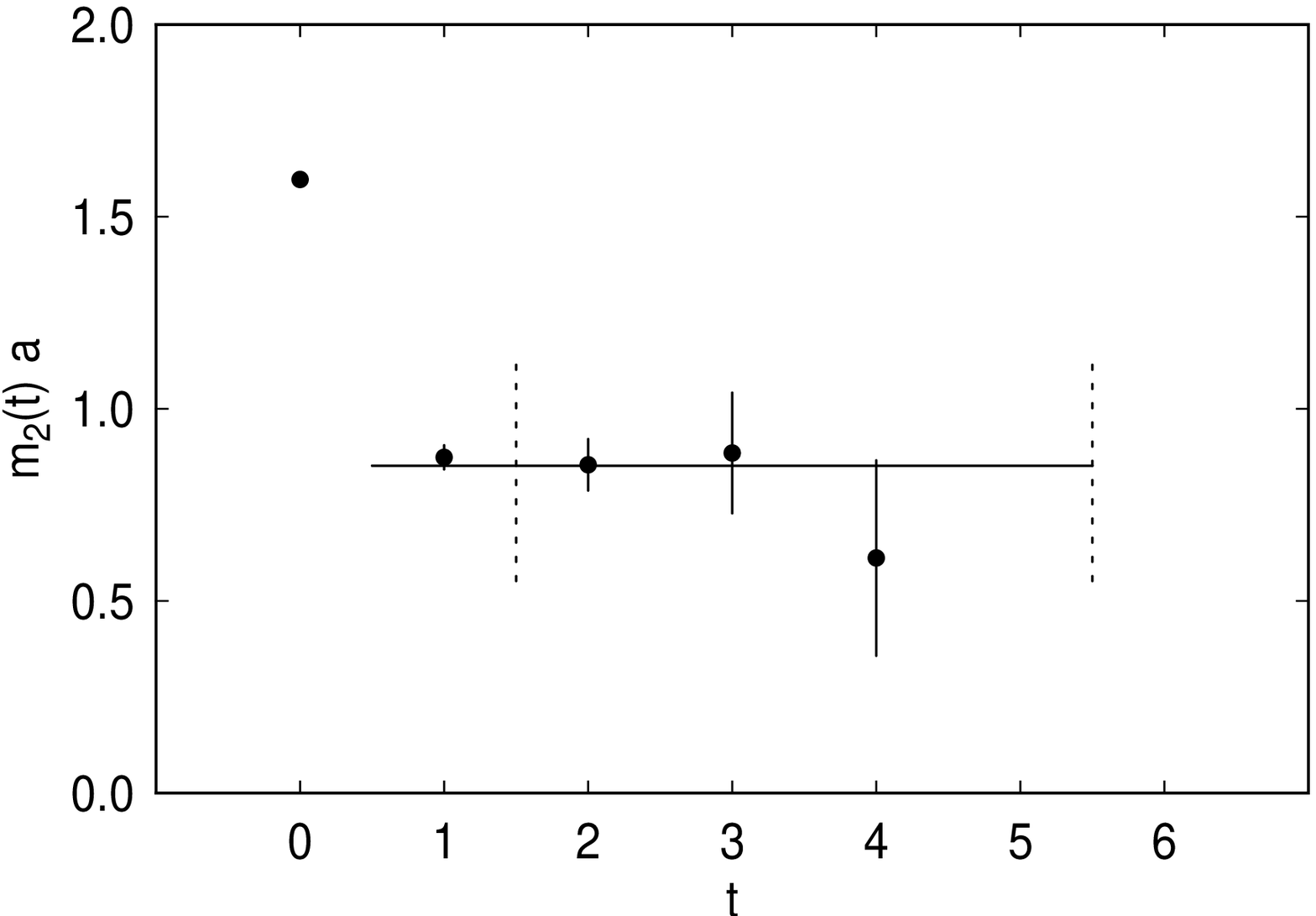}
\caption{ Effective masses and fitted mass for the
$2^{++}$ glueball with smearing size 5 on a
$30 \times 32^2 \times 40$ lattice at $\beta = 6.17$.}
\label{fig:m2at61}
\end{figure}

\begin{figure}
\epsfxsize=\textwidth \epsfbox{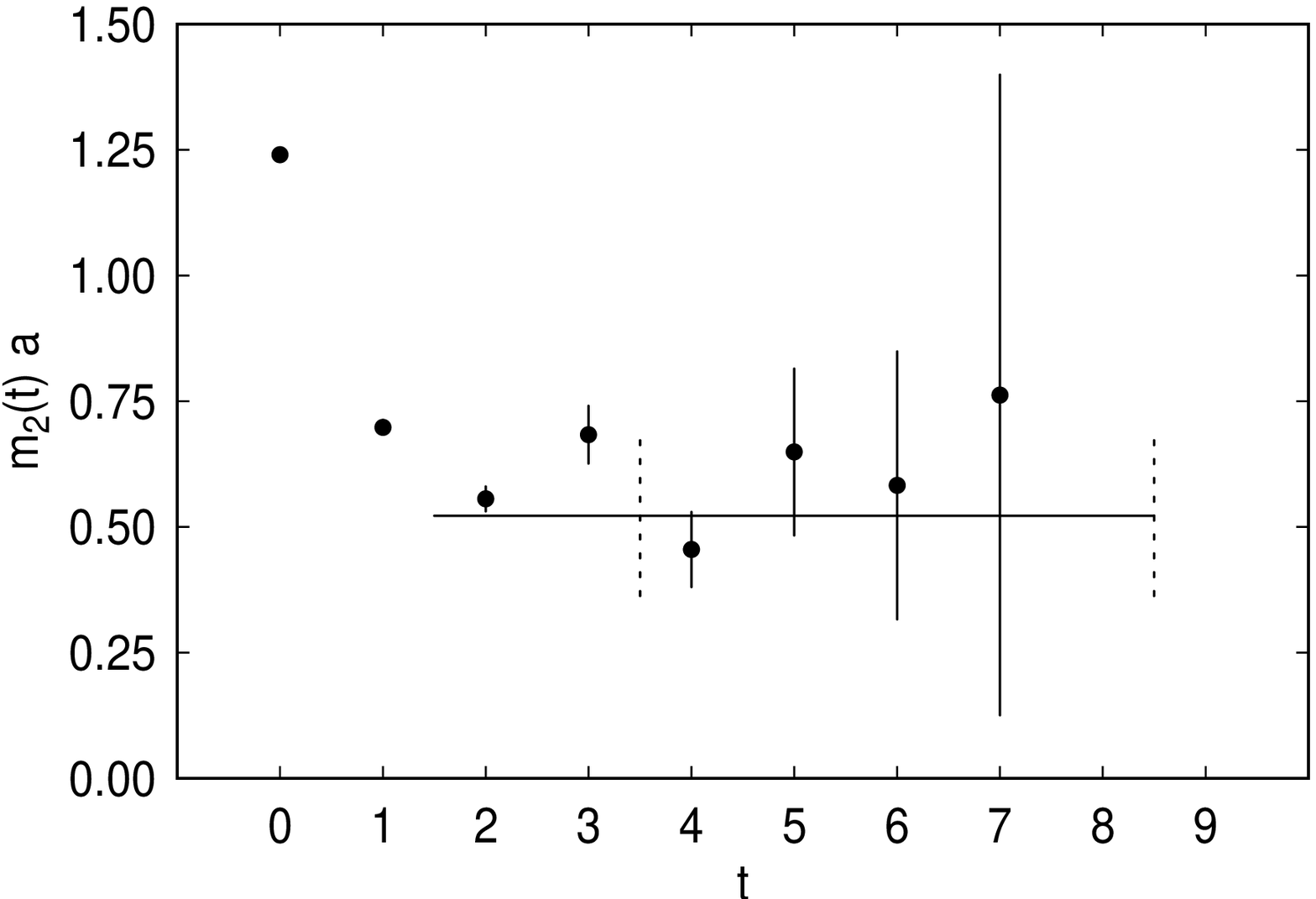}
\caption{ Effective masses and fitted mass for the $2^{++}$ 
glueball with smearing size 5 on a
$30 \times 32^2 \times 40$ lattice at $\beta = 6.40$.}
\label{fig:m2at64}
\end{figure}

To determine the range of smearing sizes and time separations from which
to extract masses for each lattice and spin, we evaluated effective
masses $m_i^s( t)$ by fitting propagators $C_i^{ss}(t)$ to
Eqs.~(\ref{defasym}) at separations $t$ and $t + 1$.  In all cases, the
time direction period $T$ was large enough and the time separation at
which our mass signals disappeared into noise small enough that the
values we found for $m_i^s( t)$ were equivalent to using the simpler
definition $m_i^s( t) a = \ln [ C_i^{ss}(t)/ C_i^{ss}(t + 1)],$ where
$a$ is the lattice spacing.  At large $t$ each $m_i^s( t)$ approaches
the corresponding $m_i$ of Eqs.~(\ref{defasym}).  It follows from the
convexity of $C_i^{ss}(t)$ that this approach will be from above.  A
plateau at large $t$ in a graph of $m_i^s( t)$ identifies the region
over which it is reasonable to try fitting propagator data to
Eqs.~(\ref{defasym}).  Figures (\ref{fig:m0at57}) - (\ref{fig:m2at64})
show effective masses for the $0^{++}$ glueball on all lattices of
Table~\ref{tab:lattices} and for the $2^{++}$ glueball on the three
lattices at largest $\beta$.  At time separations beyond those shown, we
were unable to obtain statistically significant effective mass values.
At the two lowest $\beta$ we were unable to obtain enough statistically
significant effective mass values for the $2^{++}$ glueball to find a
plateau in the mass.  In each case, we show the smearing size which
gives the best signal.  All of the data for the $0^{++}$ glueball shows
fairly clear effective mass plateaus extending over at least 4 time
values. The data for the $2^{++}$ also shows effective mass plateaus,
but not as clearly as the $0^{++}$ data.

Having found effective mass plateaus for each smearing size, we fit the
largest set of time slices which could be considered to be in the
plateau to Eqs.~(\ref{defasym}) and determined the $\chi^2$ per degree
of freedom of the best fit.  We then removed a time slice at small $t$
and repeated the fit.  In most cases, either the value of $\chi^2$ per
degree of freedom of the best fit decreased or the fitted mass decreased
by a statistically significant amount, showing that the narrower fitting
range provided a better estimate of the lowest mass.  We repeated the
process as long as either the effective mass continued to decrease
significantly or $\chi^2$ per degree of freedom decreased.  The process
was stopped when we found either a large increase in $\chi^2$ per degree
of freedom or a large increase in the fitted mass's statistical
uncertainty.  We adopted the time window before either a large increase
in $\chi^2$ or a large increase in statistical uncertainty as the best
fitting range.  In about half of the cases we considered, a fit to the
window with one fewer time slice than the best choice showed a larger
value of mass than the best choice. The convexity of the propagators
with the same smearing size at both ends implies this increase can only
occur as a result of statistical noise. The occurrence of an increase in
fitted mass about half of the time suggests that, within statistical
errors, we obtained the correct lowest mass in each case.  Tables
(\ref{tab:ranges0}) and (\ref{tab:ranges2}) show the masses and values
of $\chi^2$ per degree of freedom obtained in the process of choosing
fitting ranges for the data in Figures (\ref{fig:m0at57}) -
(\ref{fig:m2at64}).  For each sequence of fits in Tables
(\ref{tab:ranges0}) and (\ref{tab:ranges2}), the fitting range next to
last was selected as the best choice.  The vertical lines in Figures
(\ref{fig:m0at57}) - (\ref{fig:m2at64}) show the final fitting range and
the horizontal lines show the mass found for this range.  In some cases
the fitting range extends beyond the last time slice for which an
effective mass is given since, by convention, we assign each effective
mass to the smaller of the two time values from which it has been
determined.  Tables (\ref{tab:massfits0a}), (\ref{tab:massfits0b}), and
(\ref{tab:massfits2}) show the masses found by fitting, on the best
fitting ranges, propagators with the same size smearing at both ends.
For each $\beta$ and spin, the optimal fitting range determined
according to our criteria turned out to the be same for each smearing
size for which we had a reliable signal. This was not put in as a
constraint, is slightly surprising, and is probably a statistical
accident.

Final mass values for each particle were determined by making a combined
fit, as described earlier, to data for the full set of different
smearing sizes for which we had reliable masses.  The results of the
final fits are shown in Tables (\ref{tab:final0}) and
(\ref{tab:final2}).  For the $2^{++}$ glueball, at $\beta$ of 6.17, the
combined data set dictated by our selection rule consisted of smearing
sizes 4, 5 and 6 over the range from time separation 2 to time
separation 5. This fit gave an unacceptable $\chi^2$ per degree of
freedom of 10.6.  Removing the data at time separation 2 and fitting
only from separations 3 to 5 reduces $\chi^2$ of 1.2 and gives the mass
shown in Table (\ref{tab:final2}).

In addition to our calculations of propagators and masses for the
$0^{++}$ and $2^{++}$ glueballs, we also calculated propagators for the
$1^{+-}$ glueball. The effective mass plateaus in these propagators were
weak, however, and the masses which could be extracted were close to the
expected masses for torelon pairs. Thus the reliability of the predicted
$1^{+-}$ masses was not clear, and we do not report these results here.

\section{INFINITE VOLUME, CONTINUUM LIMITS}

The quantity $z = m_{0++} L$, where $L$ is the lattice period, for the
masses in Table (\ref{tab:final0}) ranges from 12.45, using the smaller
period of 30 on the lattice $30 \times 32^2 \times 40$ at $\beta$ of
6.4, to 19.46, on the lattice $24^3 \times 36$ at $\beta$ of 5.93.
Masses calculated in a large finite volume approach their values in
infinite volume according to \cite{Luscher86}
\begin{eqnarray}
m_i( z) = m_i( \infty)[ 1 - g_i \frac{ exp( \frac{-\sqrt{3}}{ 2} z)}{z}].
\end{eqnarray}
For $m_0$ the coefficient $g_0$ has been estimated \cite{Schier} to be
$190 \pm 70$. Thus our values for $m_0$ should differ from their
infinite volume values by less than 0.05\%. For $m_2$ it appears from
data in Ref.~\cite{Schier} that $g_2 < 2000$.  The range of $z$ we use
then gives $m_2$ which agree with infinite volume values to better than
0.4\%. Thus all of the masses shown in Tables (\ref{tab:final0}) and
(\ref{tab:final2}) differ from their values in infinite volume by
amounts which are negligible in comparision to the statistical
uncertainties.

\begin{figure}
\epsfxsize=\textwidth \epsfbox{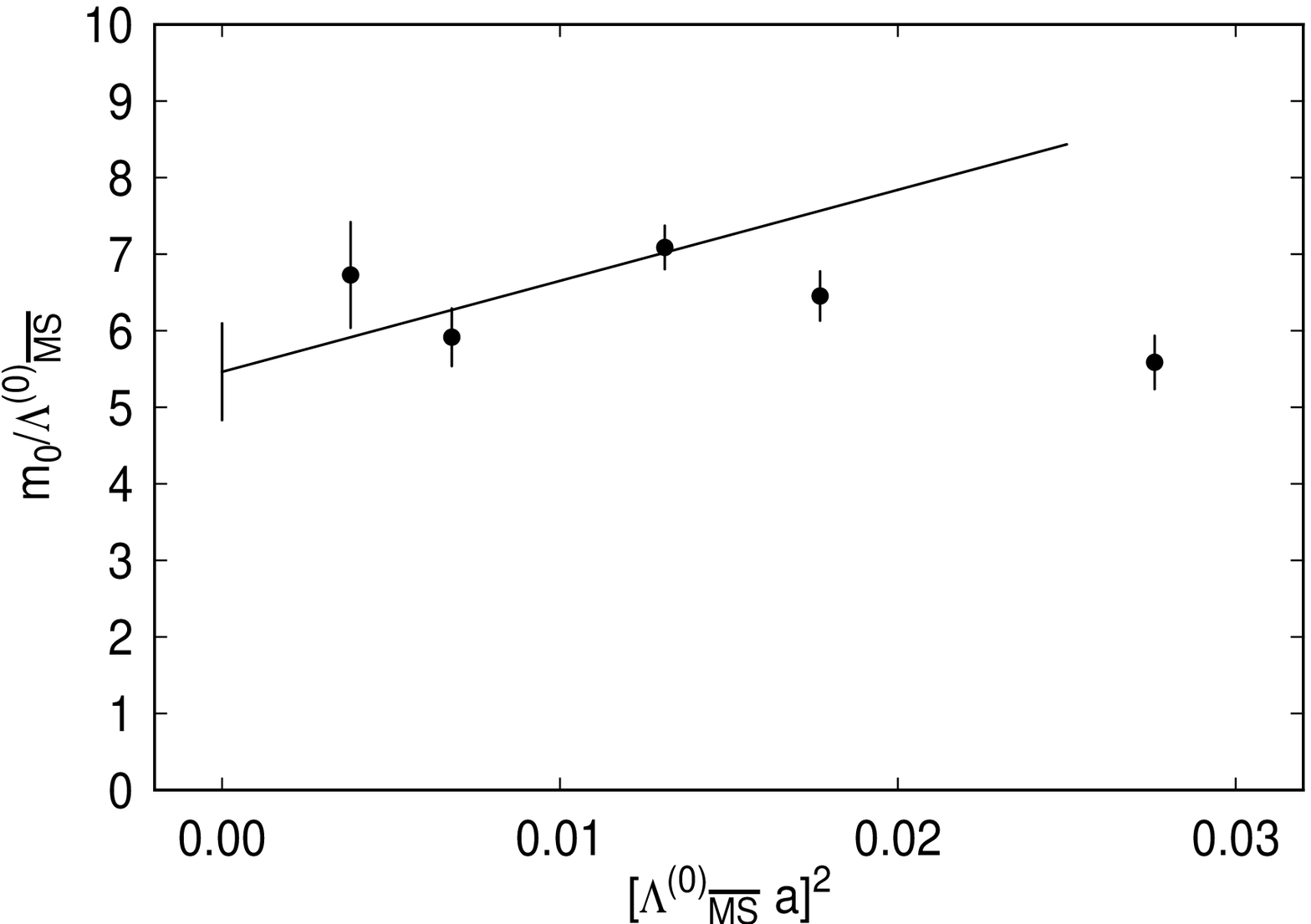}
\caption{ Values of $m_0 / \Lambda^{(0)}_{\overline{MS}}$ as a function of
$[\Lambda^{(0)}_{\overline{MS}} a]^2$ and linear extrapolation to $a = 0$.}
\label{fig:continuum0}
\end{figure}

\begin{figure}
\epsfxsize=\textwidth \epsfbox{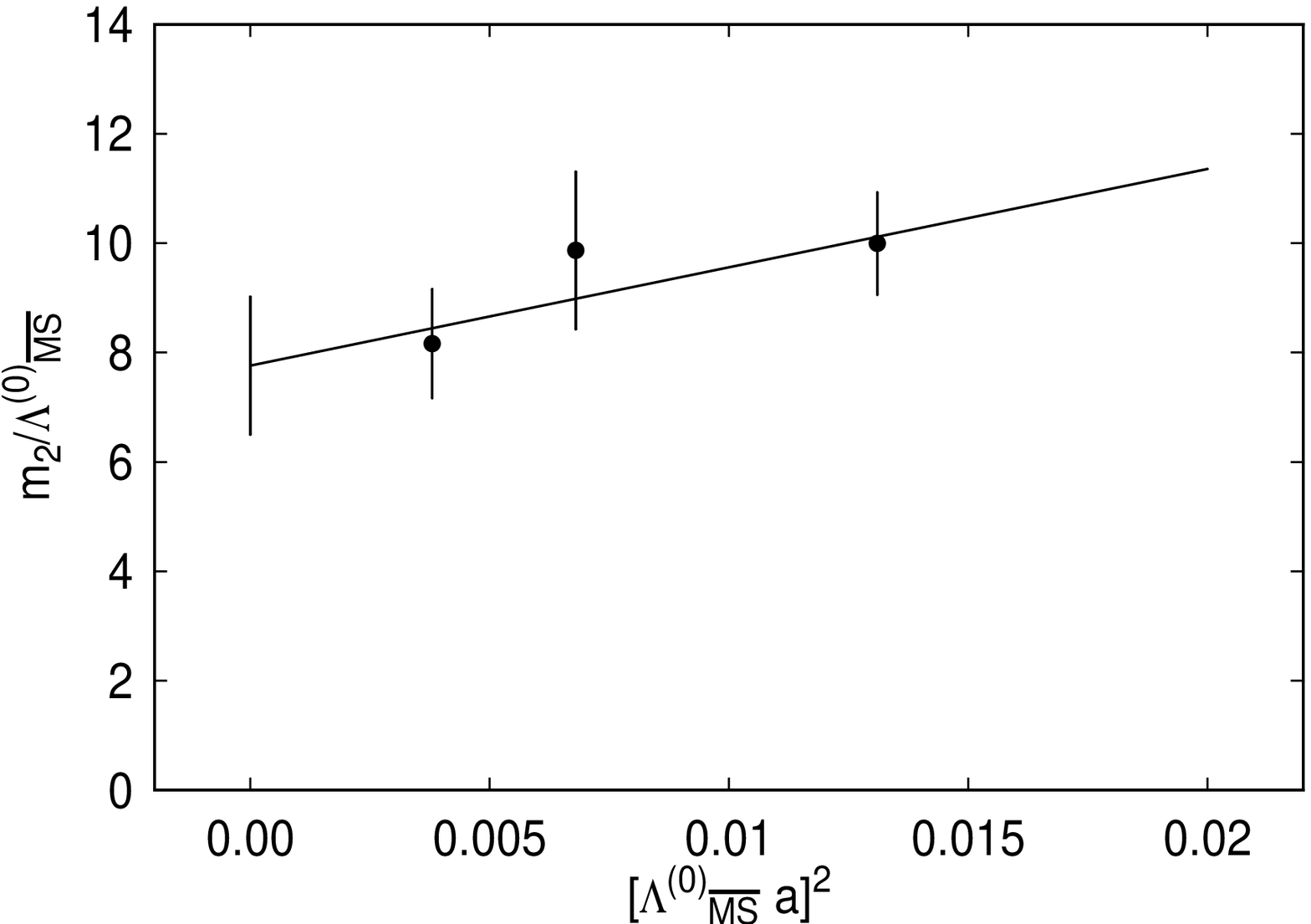}
\caption{Values of $m_2 / \Lambda^{(0)}_{\overline{MS}} $ as a function of
$[\Lambda^{(0)}_{\overline{MS}} a]^2$ and linear extrapolation to $a = 0$.}
\label{fig:continuum2}
\end{figure}

In preparation for evaluating the continuum limit of $m_0$ and $m_2$ in
physical units, the values we calculated for $m_i a$ in lattice units
were rescaled by $\Lambda^{(0)}_{\overline{MS}} a$ in lattice units.
According to mean-field improved lattice perturbation theory
\cite{Lepage,Fermilab}
\begin{eqnarray}
\label{meanfield}
\frac{1}{g^2_{\overline{MS}}} = \frac{< Tr U / 3>}{g^2} + 0.025,
\end{eqnarray}
where $g^2 = 6 / \beta$ is the lattice coupling constant, and $< Tr U >$
is the average value of a plaquette. Then with $\alpha_{\overline{MS}} =
g^2_{\overline{MS}} / 4 \pi $, the solution to the two loop
Callan-Symanzik equation for $\alpha_{\overline{MS}}( \pi / a)$ with
zero flavors of quark vacuum polarization gives
\begin{eqnarray}
\label{scaling}
\Lambda^{(0)}_{\overline{MS}} a & = & \pi 
(\frac{ b_1 \alpha_{\overline{MS}}}{ 1 + b_1
\alpha_{\overline{MS}}})^{-\frac{b_1}{b_0}} exp( -\frac{1}{b_0
\alpha_{\overline{MS}}}), \nonumber \\
b_0 & = & \frac{11}{2 \pi}, \\
b_1 & = & \frac{51}{22 \pi}. \nonumber
\end{eqnarray}
Values of $\alpha_{\overline{MS}}$ and $\Lambda^{(0)}_{\overline{MS}} a$
determined from Eqs.~(\ref{meanfield}) and (\ref{scaling}) are given in
Table (\ref{tab:lambda}).  Values of $m_i /
\Lambda^{(0)}_{\overline{MS}} $ determined from Tables
(\ref{tab:final0}), (\ref{tab:final2}) and (\ref{tab:lambda}) are shown
in Figures (\ref{fig:continuum0}) and (\ref{fig:continuum2}). In
Ref.~\cite{Butler93} it was found that the rho mass ratio $m_{\rho} /
\Lambda^{(0)}_{\overline{MS}}$ is nearly independent of $\beta$ and
within 2\% of its continuum limit once $\beta \ge 5.7$.  Thus graphs of
our glueball masses scaled by the rho mass, $m_i / m_{\rho}$, would
differ from Figures (\ref{fig:continuum0}) and (\ref{fig:continuum2})
only by a constant rescaling of the vertical axis.

Since the leading irrelevant term in the Wilson action for pure
gauge fields is $O(a^2)$, we expect that for sufficiently small lattice
spacing mass ratios will approach their continuum limits according to
\begin{eqnarray}
\label{contlim}
\frac{m_i}{\Lambda^{(0)}_{\overline{MS}}}( a) = 
\frac{m_i}{\Lambda^{(0)}_{\overline{MS}}}( 0) + c_i a^2,
\end{eqnarray}
with $a$-independent coefficients $c_i$.  Fits of $m_0$ to
Eq.~(\ref{contlim}) using five data points, using the four data points
with the smallest $a$ and using the three data points with the smallest
$a$ are listed in Table (\ref{tab:extrap}).  Since Eq.~(\ref{contlim})
is an asymptotic form approached at small lattice spacing $a$, and since
the $\chi^2$ per degree of freedom of fits to Eq.~(\ref{contlim})
progressively decreases as data points with larger $a$ are eliminated,
it appears that the fit using three data points gives the most reliable
extrapolation to the continuum limit.  It is interesting to notice that
a quadratic fit to all five data points gives essentially the same
continuum limit and same $\chi^2$ per degree of freedom as the linear
fit at the three smallest points.  For $m_2$ we have data at only the
three smallest values of $a$. A fit of Eq.~(\ref{contlim}) to these,
giving a fairly small $\chi^2$ is also shown in Table
(\ref{tab:extrap}).  The three point fits for $m_0$ and $m_2$ are shown
in Figures (\ref{fig:continuum0}) and (\ref{fig:continuum2}),
respectively. The vertical lines at $a = 0$ in each figure are the
statistical uncertainties in the extrapolated mass values.

To obtain masses in physical units from the continuum limit ratios 
$m_i / \Lambda^{(0)}_{\overline{MS}}$, we use the continum value of $245.0
\pm 9.2$ MeV for $\Lambda^{(0)}_{\overline{MS}}$ found in
Ref.~\cite{Butler93} from the continuum limit of $m_{\rho} /
\Lambda^{(0)}_{\overline{MS}}$ combined with the observed value of
$m_{\rho}$.  For $m_0$ we get $1340 \pm 160$ MeV, and for $m_2$
we find 
$1900 \pm 320$ MeV, with statistical uncertainties combining the
uncertainties in $m_i / \Lambda^{(0)}_{\overline{MS}}$ and in 
$\Lambda^{(0)}_{\overline{MS}}$.

An alternate way to find the continuum limit of lattice values of $m_i
a$ is to divide each by the square root of the string tension in lattice
units, $\sqrt{K} a$, and extrapolate $ m_i /\sqrt{K}$ to its continuum
limit. The continuum limit of $\sqrt{K}$, however, is not a directly
observable quantity. Its continuum value can be determined by
extrapolating $\sqrt{K} / m_{\rho}$ to its continuum limit, then using
the observed value of $m_{\rho}$. The resulting continuum prediction for
$\sqrt{K}$ inherits statistical errors both from lattice calculations of
$\sqrt{K}$ and from lattice calculations of $m_{\rho}$.  The continuum
value which we have used for $\Lambda^{(0)}_{\overline{MS}}$, on the
other hand, has statistical errors only from lattice calculations of
$m_{\rho}$.  Thus the method we have adopted should give a smaller
uncertainty for continuum mass predictions in physical units. The
central value of the answers obtained extrapolating $m_i / \sqrt{K}$ do
not appear to be significantly different from our present answers.  For
example, interpolating published values of $\sqrt{K} a$ \cite{Tension}
to our values of $\beta$, using these to find the continuum limit of
$m_0 / \sqrt{K}$, and assuming the popular value of 440 MeV for
continuum $\sqrt{K}$, gives a continuum $m_0$ of about 1310 MeV.

As mentioned in Sect.~(\ref{intro}), continuum limit predictions for
$m_0$ of $1550 \pm 50$ MeV and for $m_2$ of $2270 \pm 100$ MeV were
reported recently in Ref.~\cite{Bali93}. These numbers were obtained by
taking $m_0 / \sqrt{K}$ and $m_2 / \sqrt{K}$ at $\beta$ of 6.4 as
continuum values. If the fit in Ref.~\cite{Bali93} of $m_0 / \sqrt{K}$
to a linear function of $K a^2$ is actually extrapolated to the
continuum, the central value of $m_0$ becomes 1600 MeV. This number
differs from our result by about 1.6 standard deviations.  It appears to
us, however, that the central values of both predictions quoted in
Ref.~\cite{Bali93} may be somewhat higher and the error bars on both
predictions somewhat smaller than the data of Ref.~\cite{Bali93} itself
suggests.

For $m_0$ and $m_2$ at $\beta$ of 6.4, Ref.~\cite{Bali93} takes the
effective masses found from the $0^{++}$ and $2^{++}$ propagators
between time separations 2 and 3.  From the effective mass tables given
in Ref.~\cite{Bali93} it seems unclear how close the effective masses at
these time separations are to their asymptotic large time separation
values and how reliable the error estimates on these effective masses
may be.  The central values of effective masses found from the $0^{++}$
and $2^{++}$ propagators fall monotonically at time separations larger
than 2.  Both effective masses between time separations of 4 and 5, for
example, are about 10\% lower than corresponding numbers between
separations 2 and 3 and have error bars which are larger than the
corresponding errors by factors of more than 2.  It also seems to us
uncertain whether the $m_0$ value which Ref.~\cite{Bali93} takes from
Ref.~\cite{Michael88} for $\beta$ of 6.2 is a reliable estimate of the
asymptotic effective mass at large time separation. Our calculation of
$m_0$ at $\beta$ of 6.17, which should lie above $m_0$ at 6.2, is
actually lower by about 10\%.  The data of Ref.~\cite{Michael88} for
$m_0$ at $\beta$ of 5.9 is consistent with our fit of $m_0$ to a linear
function of $[\Lambda^{(0)}_{\overline{MS}} a]^2$, while $m_2$ is again
above the extrapolation of our data.

The results of Refs.~\cite{Bali93} and \cite{Michael88} become closer to
ours if $m_0$ and $m_2$ are taken from effective masses at larger time
separations approximating the ranges over which our mass fits were done.
Combining effective masses between time separations 2 and 3 at $\beta$
of 5.9 and 6.0 \cite{Michael88}, time separations 3 and 4 at $\beta$ of
6.2 \cite{Michael88}, and time separations 4 and 5 at $\beta$ of 6.4
\cite{Bali93}, we fit $m_i / \Lambda^{(0)}_{\overline{MS}}$ to linear
functions of $[\Lambda^{(0)}_{\overline{MS}} a]^2$. 
The continuum $m_0 /
\Lambda^{(0)}_{\overline{MS}}$ becomes $6.11 \pm 0.52$ and $m_2 /
\Lambda^{(0)}_{\overline{MS}}$ becomes $7.75 \pm 0.89$.  With
$\Lambda^{(0)}_{\overline{MS}}$ of $245.0
\pm 9.2$ MeV, the continuum prediction for $m_0$ is $1500 \pm 140$
MeV and for $m_2$ is $1900 \pm 230$ MeV. These results are both less
than one standard deviation from our corresponding predictions.  A fit
to this set of data combined with ours in the region $\beta \ge 5.9$
gives a continuum limit for $m_0 / \Lambda^{(0)}_{\overline{MS}}$ of
$5.89 \pm 0.38$ and for $m_2 /
\Lambda^{(0)}_{\overline{MS}}$ of $7.77 \pm 0.73$. These values
yield $m_0$ of $1440 \pm 110$ MeV and $m_2$ of $1900 \pm 190$
MeV.

We would like to thank Frank Butler for writing some of the analysis
software which we used, and Mike Cassera, Molly Elliott, Dave George,
Chi Chai Huang and Ed Nowicki for their work on GF11.

\newpage

\begin{table}
\begin{center}
\begin{tabular}{|r|c|c|c|}     \hline
 lattice & $\beta$ & skip & count \\ \hline
 $16^3 \times 24$ & 5.70 & 400 & 4039\\ 
 $20^3 \times 30$ & 5.83 & 400 & 4002\\ 
 $24^3 \times 36$ & 5.93 & 400 & 4004\\
 $30 \times 32^2 \times 40$ & 6.17 & 400 & 2005\\
 $30 \times 32^2 \times 40$ & 6.40 & 400 & 2002\\
 \hline
\end{tabular}
\caption{Configurations analyzed.}
\label{tab:lattices}
\end{center}
\end{table}

\begin{table}
\begin{center}
\begin{tabular}{|c|c|c|c|c|}     \hline
$\beta$ & size & range & mass & $\chi^2$ \\ \hline

5.70 & 2 & 1-5 & $0.958 \pm 0.023$ & 0.56 \\ 
5.70 & 2 & 2-5 & $0.896 \pm 0.058$ & 0.20 \\  
5.70 & 2 & 3-5 & $0.978 \pm 0.176$ & 0.03 \\ \hline 
5.83 & 2 & 1-5 & $0.896 \pm 0.017$ & 0.93 \\ 
5.83 & 2 & 2-5 & $0.845 \pm 0.039$ & 0.62 \\  
5.83 & 2 & 3-5 & $0.798 \pm 0.076$ & 0.92 \\ \hline 
5.93 & 2 & 1-5 & $0.856 \pm 0.015$ & 1.15 \\
5.93 & 2 & 2-5 & $0.833 \pm 0.033$ & 1.54 \\ 
5.93 & 2 & 3-5 & $0.871 \pm 0.085$ & 2.78 \\ \hline
6.17 & 4 & 2-7 & $0.564 \pm 0.028$ & 1.55 \\ 
6.17 & 4 & 3-7 & $0.482 \pm 0.035$ & 0.20 \\  
6.17 & 4 & 4-7 & $0.461 \pm 0.058$ & 0.20 \\ \hline
6.40 & 5 & 3-9 & $0.444 \pm 0.029$ & 1.12 \\ 
6.40 & 5 & 4-9 & $0.387 \pm 0.035$ & 0.34 \\ 
6.40 & 5 & 5-9 & $0.399 \pm 0.050$ & 0.43 \\ \hline 
\end{tabular}
\end{center}
\caption{Masses, in units of $1/a$, obtained for $0^{++}$ glueballs over 
various fitting ranges.}
\label{tab:ranges0}
\end{table}

\begin{table}
\begin{center}
\begin{tabular}{|c|c|c|c|c|c|}     \hline
$\beta$ & size & range & mass & $\chi^2$ \\ \hline

5.93 & 4 & 1-4 & $1.223 \pm 0.050$ & 0.03 \\  
5.93 & 4 & 2-4 & $1.197 \pm 0.188$ & 0.03 \\  
5.93 & 4 & 3-4 & $1.288 \pm 0.766$ &      \\ \hline 
6.17 & 5 & 1-5 & $0.871 \pm 0.029$ & 0.22 \\  
6.17 & 5 & 2-5 & $0.852 \pm 0.061$ & 0.28 \\ 
6.17 & 5 & 3-5 & $0.835 \pm 0.137$ & 0.55 \\ \hline
6.40 & 5 & 2-8 & $0.579 \pm 0.023$ & 1.25 \\ 
6.40 & 5 & 3-8 & $0.632 \pm 0.053$ & 1.09 \\ 
6.40 & 5 & 4-8 & $0.522 \pm 0.059$ & 0.49 \\
6.40 & 5 & 5-8 & $0.641 \pm 0.128$ & 0.04 \\ \hline 
\end{tabular}
\end{center}
\caption{Masses, in units of $1/a$, obtained for $2^{++}$ glueballs over 
various fitting ranges.}
\label{tab:ranges2}
\end{table}

\begin{table}
\begin{center}
\begin{tabular}{|c|c|c|c|c|c|}     \hline
$\beta$ & size & range & mass & $\chi^2$ \\ \hline

5.70 & 0 & 2-5 & $0.985 \pm 0.141$ & 0.01  \\ 
5.70 & 1 & 2-5 & $0.948 \pm 0.064$ & 0.54  \\ 
5.70 & 2 & 2-5 & $0.896 \pm 0.058$ & 0.20  \\ 
5.70 & 3 & 2-5 & $0.842 \pm 0.072$ & 0.57  \\ 
5.70 & 4 & 2-5 & $0.797 \pm 0.103$ & 0.87  \\ \hline 
5.83 & 1 & 2-5 & $0.889 \pm 0.056$ & 2.02  \\
5.83 & 2 & 2-5 & $0.845 \pm 0.039$ & 0.62  \\ 
5.83 & 3 & 2-5 & $0.840 \pm 0.047$ & 1.30  \\ 
5.83 & 4 & 2-5 & $0.845 \pm 0.075$ & 1.36  \\ 
5.83 & 5 & 2-5 & $0.834 \pm 0.116$ & 0.79  \\ 
5.83 & 6 & 2-5 & $0.798 \pm 0.105$ & 1.00  \\ \hline 
\end{tabular}
\caption{$0^{++}$ glueball masses, in units of $1/a$, fitted to optimal 
ranges for each size of smearing.}
\label{tab:massfits0a}
\end{center}
\end{table}

\begin{table}
\begin{center}
\begin{tabular}{|c|c|c|c|c|c|}     \hline
$\beta$ & size & range & mass & $\chi^2$ \\ \hline

5.93 & 1 & 2-5 & $0.905 \pm 0.053$ & 0.13  \\ 
5.93 & 2 & 2-5 & $0.833 \pm 0.031$ & 1.54  \\ 
5.93 & 3 & 2-5 & $0.796 \pm 0.030$ & 3.19  \\
5.93 & 4 & 2-5 & $0.781 \pm 0.043$ & 3.98  \\ 
5.93 & 5 & 2-5 & $0.746 \pm 0.060$ & 2.41  \\ 
5.93 & 6 & 2-5 & $0.724 \pm 0.090$ & 1.52  \\ \hline 
6.17 & 2 & 3-7 & $0.504 \pm 0.055$ & 0.83  \\ 
6.17 & 3 & 3-7 & $0.505 \pm 0.035$ & 0.37  \\ 
6.17 & 4 & 3-7 & $0.482 \pm 0.035$ & 0.20  \\ 
6.17 & 5 & 3-7 & $0.473 \pm 0.041$ & 0.04  \\ 
6.17 & 6 & 3-7 & $0.495 \pm 0.060$ & 0.30  \\ \hline 
6.40 & 3 & 4-9 & $0.392 \pm 0.043$ & 0.38  \\
6.40 & 4 & 4-9 & $0.389 \pm 0.037$ & 0.27  \\
6.40 & 5 & 4-9 & $0.387 \pm 0.035$ & 0.34  \\ 
6.40 & 6 & 4-9 & $0.392 \pm 0.034$ & 0.52  \\ 
6.40 & 7 & 4-9 & $0.400 \pm 0.036$ & 0.66  \\ 
6.40 & 8 & 4-9 & $0.408 \pm 0.042$ & 0.77  \\ \hline 
\end{tabular}    
\end{center}
\caption{$0^{++}$ glueball masses, in units of $1/a$, fitted to optimal 
ranges for each size of smearing.}
\label{tab:massfits0b}
\end{table}

\begin{table}
\begin{center}
\begin{tabular}{|c|c|c|c|c|c|}     \hline
$\beta$ & size & range & mass & $\chi^2$ \\ \hline

5.93 & 3 & 2-4 & $1.131 \pm 0.109$ & 0.11 \\ 
5.93 & 4 & 2-4 & $1.197 \pm 0.188$ & 0.03 \\ 
5.93 & 5 & 2-4 & $1.464 \pm 0.405$ & 0.19 \\ \hline 
6.17 & 4 & 2-5 & $0.842 \pm 0.059$ & 0.18 \\
6.17 & 5 & 2-5 & $0.852 \pm 0.062$ & 0.28 \\ 
6.17 & 6 & 2-5 & $0.887 \pm 0.080$ & 0.22 \\ \hline 
6.40 & 4 & 4-8 & $0.514 \pm 0.066$ & 0.41 \\
6.40 & 5 & 4-8 & $0.522 \pm 0.059$ & 0.49 \\ 
6.40 & 6 & 4-8 & $0.510 \pm 0.062$ & 0.81 \\ \hline
\end{tabular}
\end{center}
\caption{$2^{++}$ glueball masses, in units of $1/a$, fitted to optimal 
ranges for each size of smearing.}
\label{tab:massfits2}
\end{table}

\begin{table}
\begin{center}
\begin{tabular}{|c|c|c|c|c|c|}     \hline
$\beta$ & sizes & range & mass & $\chi^2$ \\ \hline
5.70 & 0-4 & 2-5 & $0.928 \pm 0.058$ & 1.35 \\ \hline
5.83 & 1-6 & 2-5 & $0.858 \pm 0.043$ & 1.80 \\ \hline  
5.93 & 1-6 & 2-5 & $0.811 \pm 0.033$ & 2.70 \\ \hline 
6.17 & 2-6 & 3-7 & $0.489 \pm 0.031$ & 1.29 \\ \hline 
6.40 & 3-8 & 4-9 & $0.415 \pm 0.043$ & 1.68 \\ \hline 
\end{tabular}
\end{center}
\caption{$0^{++}$ glueball masses, in units of $1/a$, obtained from 
fits to propagator matrices for sets of several smearing sizes.}
\label{tab:final0}
\end{table}

\begin{table}
\begin{center}
\begin{tabular}{|c|c|c|c|c|c|}     \hline
$\beta$ & sizes & range & mass & $\chi^2$ \\ \hline
5.93 &  3-5 & 2-4 & $1.144 \pm 0.107$ & 2.58 \\ \hline 
6.17 &  4-6 & 3-5 & $0.816 \pm 0.119$ & 1.17 \\ \hline 
6.40 &  4-6 & 4-8 & $0.504 \pm 0.061$ & 2.22 \\ \hline 
\end{tabular}
\end{center}
\caption{$2^{++}$ glueball masses, in units of $1/a$, obtained from 
fits to propagator matrices for sets of several smearing sizes.}
\label{tab:final2}
\end{table}

\begin{table}
\begin{center}
\begin{tabular}{|r|c|c|c|}     \hline
 lattice & $\beta$ & $\alpha_{\overline{MS}}$ & 
 $\Lambda_{\overline{MS}} a$ \\ \hline
 $16^3 \times 24$ & 5.70 & 0.1456 & 0.1661\\ 
 $20^3 \times 30$ & 5.83 & 0.1370 & 0.1329\\ 
 $24^3 \times 36$ & 5.93 & 0.1318 & 0.1144\\
 $30 \times 32^2 \times 40$ & 6.17 & 0.1218 & 0.08265\\
 $30 \times 32^2 \times 40$ & 6.40 & 0.1141 & 0.06177\\
 \hline
\end{tabular}
\end{center}
\caption{Values of $\alpha_{\overline{MS}}$ and $\Lambda_{\overline{MS}} a$.
Statistical uncertainties are all smaller than 1 in the last decimal
place.}
\label{tab:lambda}
\end{table}

\begin{table}
\begin{center}
\begin{tabular}{|c|c|c|c|c|}     \hline
 $J^{PC}$ & fit & points & $m_i / \Lambda_{\overline{MS}}$ 
& $\chi^2$ \\ \hline
$0^{++}$ & linear    & 3 & $5.46 \pm 0.64$  & 2.36 \\
$0^{++}$ & linear    & 4 & $6.19 \pm 0.50$  & 2.89 \\
$0^{++}$ & linear    & 5 & $6.97 \pm 0.29$  & 3.43 \\
$0^{++}$ & quadratic & 5 & $5.37 \pm 0.81$  & 2.26 \\
$2^{++}$ & linear    & 3 & $7.76 \pm 1.26$  & 0.47 \\ \hline
\end{tabular}
\end{center}
\caption{Extrapolations of $m_i / \Lambda_{\overline{MS}}$ to the
continuum limit using either linear or quadratic functions of 
$[\Lambda_{\overline{MS}} a]^2$.}
\label{tab:extrap}
\end{table}

\newpage

\clearpage

\end{document}